\documentclass[aps,pra,showpacs,floatfix,twocolumn]{revtex4-1}
\usepackage{graphicx}
\usepackage{amsmath}
\usepackage{bm}
\usepackage{color}
\usepackage{dcolumn}
\usepackage{multirow}
\usepackage{hyperref}

\begin{document}

\title{Hyper-Ramsey spectroscopy with probe laser intensity fluctuations}

\newcommand{\NIST}{
National Institute of Standards and Technology, 325 Broadway, Boulder, Colorado 80305, USA}

\author{K. Beloy}
\affiliation{\NIST}

\date{\today}

\begin{abstract}
We examine the influence of probe laser intensity fluctuations on hyper-Ramsey spectroscopy. We assume, as is appropriate for relevant cases of interest, that the probe laser intensity $I$ determines both the Rabi frequency $(\propto\sqrt{I})$ and the frequency shift to the atomic transition $(\propto I)$ during probe laser interactions with the atom. The spectroscopic signal depends on these two quantities that co-vary with fluctuations in the probe laser intensity. Introducing a simple model for the fluctuations, we find that the signature robustness of the hyper-Ramsey method can be compromised. Taking the Yb$^+$ electric octupole clock transition as an example, we quantify the clock error under different levels of probe laser intensity fluctuations. 
\end{abstract}


\maketitle


State-of-the-art atomic frequency standards are based on optical transitions between long-lived ``clock'' states of an atom~\cite{LudBoyYe15}. Choosing sufficiently long-lived clock states ensures that the natural linewidth does not limit spectroscopic discrimination of the transition frequency. However, the clock states' inherent reluctance to decay is accompanied by the clock transition's inherent reluctance to be laser-driven. Thus, high probe laser intensity may be required to drive the clock transition on practical timescales. This high intensity, in turn, can give unwelcome prominence to the probe-induced ac Stark shift of the clock transition, which scales linearly with the intensity. This probe Stark shift must be given due consideration in order to distill the unperturbed clock frequency from spectroscopic measurements. As a striking example, consider the ${^2S_{1/2}}\rightarrow{^2F_{7/2}}$ electric octupole clock transition in Yb$^+$, where a $\pi$-pulse driven in $70$~ms or less implies an accompanying fractional clock shift of $10^{-13}$ or greater~\cite{HunOkhLip12}.

To address the problem of probe Stark shifts (more generally, probe-related shifts), Yudin {\it et al.}~\cite{YudTaiOat10} conceived a spectroscopic protocol based on composite pulses, which they dubbed hyper-Ramsey spectroscopy (HRS). The method is recapitulated below. The seminal work spawned numerous theoretical~\cite{TabTaiYud13,ZanAlmdeC14,TabTaiDmi15,ZanYudTai15,ZandeCAri16,HobBowKin16,YudTaiBas16,ZanLefTai17} and experimental~\cite{HunLipOkh12,HobBowKin16} studies, including generalizations and proof-of-principle demonstrations. Most notably, HRS was recently implemented in a single-ion clock based on the aforementioned electric octupole transition in Yb$^+$~\cite{HunSanLip16}. Relative to its predecessor based exclusively on Rabi spectroscopy~\cite{HunOkhLip12}, this clock boasts a significantly reduced probe Stark shift uncertainty, with a total fractional clock uncertainty given at $3.2\times10^{-18}$.

Here we investigate HRS in the presence of probe intensity fluctuations, assuming a frequency shift to the atomic transition proportional to the intensity (e.g., the probe Stark shift). We further acknowledge dependence of the Rabi frequency on the probe intensity, being proportional to the square-root of intensity for single-photon transitions. Thus, the frequency shift and the Rabi frequency co-vary with fluctuations in the probe intensity. Respecting this covariance is critical for a proper treatment. Clock operation involves matching a local oscillator frequency $\omega_\mathrm{LO}$ to the unperturbed atomic transition frequency $\omega_0$, with a difference $\omega_\mathrm{LO}-\omega_0$ being regarded as a clock error. The term frequency is used here and below in lieu of angular frequency for brevity.


Central to our analysis is the transition probability from the ground state $\left|g\right\rangle=\left(\begin{smallmatrix}0\\1\end{smallmatrix}\right)$ to the excited state $\left|e\right\rangle=\left(\begin{smallmatrix}1\\0\end{smallmatrix}\right)$ following the HRS interrogation sequence, which can be viewed as a sequence of rotations to the Bloch vector describing our pseudo-spin-1/2 quantum system. The rotation operator is
\begin{align}
U\left(\bm{\phi}\right)
&{}=\exp\left(-\frac{i}{2}\bm{\phi}\cdot\bm{\sigma}\right)
\nonumber\\&{}
=\mathcal{I}\cos\left(\frac{\phi}{2}\right)-i\left(\bm{\hat{\phi}}\cdot\bm{\sigma}\right)\sin\left(\frac{\phi}{2}\right),
\label{Eq:Urot}
\end{align}
where $\mathcal{I}$ is the $2\times2$ identity matrix and where Cartesian components of $\bm{\sigma}$ are the conventional Pauli spin matrices. The argument $\bm{\phi}$ specifies both the angle $\phi\equiv\left|\bm{\phi}\right|$ and axis $\bm{\hat{\phi}}\equiv\bm{\phi}/\phi$ of rotation. For a given sequence of rotations, the transition probability reads
\begin{equation}
P=\left|\left\langle e\left|
\cdots U\left(\bm{\phi}_3\right)U\left(\bm{\phi}_2\right)U\left(\bm{\phi}_1\right)
\right|g\right\rangle\right|^2,
\label{Eq:transprob}
\end{equation}
where the arguments $\bm{\phi}_i$ are labeled sequentially. Expressions (\ref{Eq:Urot}) and (\ref{Eq:transprob}) facilitate straightforward evaluation of the transition probability once the arguments $\bm{\phi}_i$ are specified.

Table \ref{Tab:sequence} specifies the arguments $\bm{\phi}_i$ for the HRS interrogation sequence. In line with previous studies, we work in the rotating frame and employ the rotating wave approximation. Steps 1, 4, and 6 represent interactions with the probe laser, having respective durations $\tau$, $2\tau$, and $\tau$. The interaction strength is quantified by the Rabi frequency $\Omega$. For optimal contrast, $\Omega=\pi/2\tau$ is desired. With this value, steps 1 and 6 correspond to ideal $\pi/2$-pulses, while step 4 corresponds to an ideal $\pi$-pulse. In practice, however, only an approximate realization of this $\Omega$ can be expected.
Accompanying each interaction is a shift to the atomic transition frequency, $\Delta_\mathrm{shift}$. In an attempt to compensate for this shift, the laser frequency is ``stepped'' by an amount $\Delta_\mathrm{step}$ relative to $\omega_\mathrm{LO}$, with $\Delta_\mathrm{step}\approx\Delta_\mathrm{shift}$. That is, during the interactions, the probe laser frequency and the atomic transition frequency are given by $\omega_\mathrm{LO}+\Delta_\mathrm{step}$ and $\omega_0+\Delta_\mathrm{shift}$, respectively. Resonant interaction is identified with an equality of these two frequencies.

\begin{table}[tb]
\caption{Interrogation sequence for HRS and Rabi spectroscopy, with arguments $\bm{\phi}_i$ to be employed in Eqs.~(\ref{Eq:Urot}) and (\ref{Eq:transprob}). For each step $i$, Cartesian components $\phi_x$ and $\phi_z$ are specified ($\times{-1}$), while $\phi_y=0$.}
\label{Tab:sequence}
\begin{ruledtabular}
\begin{tabular}{lcc}
$i$					& $-\phi_x$			& $-\phi_z$	\\
\hline\vspace{-3mm}\\
\multicolumn{3}{c}{\it hyper-Ramsey spectroscopy}\\
1.	& $\Omega\tau$		& $\left[\left(\omega_\mathrm{LO}+\Delta_\mathrm{step}\right)-\left(\omega_0+\Delta_\mathrm{shift}\right)\right]\tau$	\\
2.	& 0					& $\left(\omega_\mathrm{LO}-\omega_0\right)T$ \\	
3.	& 0					& $\pi+\ell\pi/2$ \\
4.	& $\Omega(2\tau)$	& $\left[\left(\omega_\mathrm{LO}+\Delta_\mathrm{step}\right)-\left(\omega_0+\Delta_\mathrm{shift}\right)\right](2\tau)$ \\
5.	& 0					& $\pi$ \\
6.	& $\Omega\tau$		& $\left[\left(\omega_\mathrm{LO}+\Delta_\mathrm{step}\right)-\left(\omega_0+\Delta_\mathrm{shift}\right)\right]\tau$	\\
\vspace{-1mm}\\
\multicolumn{3}{c}{\it Rabi spectroscopy}\\
1.	& $\Omega(2\tau)$		& $\left[\left(\omega_\mathrm{LO}+\Delta_\mathrm{step}+\ell\Delta_\mathrm{hop}\right)-\left(\omega_0+\Delta_\mathrm{shift}\right)\right](2\tau)$ 
\end{tabular}
\end{ruledtabular}
\end{table}


The nominal $\pi/2$-pulse that initiates the HRS interrogation sequence is followed by free evolution for a duration $T$ (step 2). The shift to the atomic transition frequency is absent, and the step to the laser frequency is removed in kind. The local oscillator frequency $\omega_\mathrm{LO}$ is identified with the laser frequency during this free evolution interval.

Finally, steps 3 and 5 of the HRS interrogation sequence represent laser phase jumps of $\pi+\ell\pi/2$ and $\pi$, respectively \cite{myfootnote}. 
The parameter $\ell$ is discussed in the following paragraph. With the exception of these explicit phase jumps, laser phase continuity is maintained throughout the sequence (e.g., at initiation or termination of the laser frequency steps).

\begin{figure}[tb]
\includegraphics[width=\linewidth]{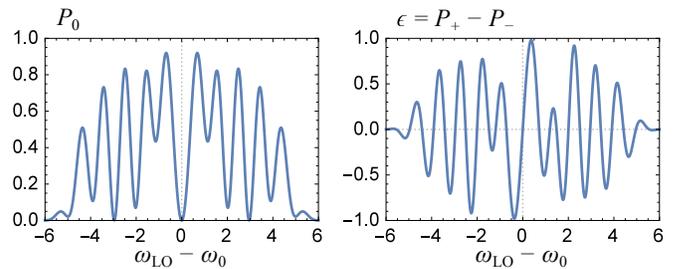}
\caption{Transition probability $P_0$ (left panel) and discriminator $\epsilon=P_+-P_-$ (right panel) for HRS with $\Delta_\mathrm{step}=\Delta_\mathrm{shift}$. A free evolution time $T=4\tau$ and Rabi frequency $\Omega=\pi/2\tau$ are assumed for these plots. The abscissas are in units of $1/\tau$.}
\label{Fig:spectra}
\end{figure}

Allowed values for $\ell$ are zero and $\pm1$. We distinguish the transition probability for these cases by subscript, $P_0$ and $P_\pm$. Taking $\ell=0$ recovers the prototypical HRS interrogation sequence. For $\Delta_\mathrm{step}=\Delta_\mathrm{shift}$, the transition probability $P_0$ amounts to a symmetric fringe pattern with respect to the detuning $\omega_\mathrm{LO}-\omega_0$, as illustrated in Fig.~\ref{Fig:spectra}. A combination of $\ell=+1$ and $\ell=-1$, on the other hand, furnishes a useful discriminator (``error'') signal,
\begin{equation}
\epsilon=P_{+}-P_{-}.
\label{Eq:errorsignal}
\end{equation}
For $\Delta_\mathrm{step}=\Delta_\mathrm{shift}$, the discriminator $\epsilon$ is antisymmetric with respect to the detuning $\omega_\mathrm{LO}-\omega_0$, as illustrated in Fig.~\ref{Fig:spectra}. Practical clock operation involves alternating between $\ell=+1$ and $\ell=-1$ interrogations and steering $\omega_\mathrm{LO}$ to realize a null value of $\epsilon$.

\begin{figure}[tb]
\includegraphics[width=0.75\linewidth]{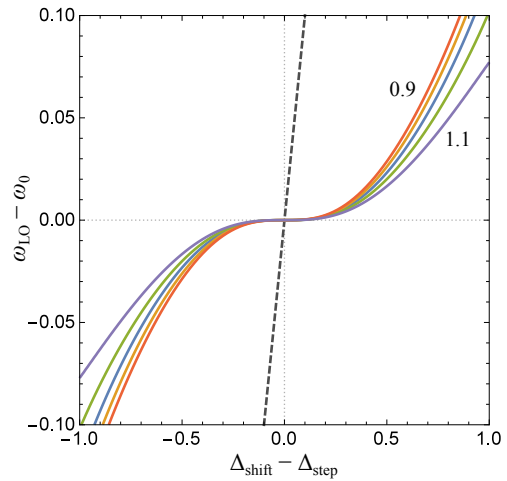}
\caption{Contours of the discriminator $\epsilon$ corresponding to a null value, $\epsilon=0$. For HRS (full curves), a free-evolution time $T=4\tau$ and Rabi frequency $\Omega=q\pi/2\tau$ are assumed, where $q$ is taken from 0.9 to 1.1 in increments of 0.05. For Rabi spectroscopy (dashed line), the contour is an $\Omega$-independent line satisfying $\omega_\mathrm{LO}-\omega_0=\Delta_\mathrm{shift}-\Delta_\mathrm{step}$. Both axes are in units of $1/\tau$.}
\label{Fig:contourstandard}
\end{figure}

The proviso $\Delta_\mathrm{step}=\Delta_\mathrm{shift}$ used in the preceding paragraph represents an experimental idealization. Introducing a misbalance $\Delta_\mathrm{step}\neq\Delta_\mathrm{shift}$ modifies the discriminator signal. Most notably, the zero-crossing of $\epsilon$ gets displaced from zero detuning. This translates to a clock error, as $\omega_\mathrm{LO}$ steers to a frequency other than $\omega_0$. This clock error is illustrated in Fig.~\ref{Fig:contourstandard}, where contours of $\epsilon$ corresponding to a null value are plotted over a range of $\left(\Delta_\mathrm{shift}-\Delta_\mathrm{step}\right)$ and $\left(\omega_\mathrm{LO}-\omega_0\right)$ values.

To provide a basis for comparison, we also consider elementary Rabi spectroscopy. For proper correspondence, a laser frequency step $\Delta_\mathrm{step}$ is included and the duration of the interaction is taken to be $2\tau$. We supplement $\Delta_\mathrm{step}$ with an additional frequency step $\ell\Delta_\mathrm{hop}$ to acquire a useful discriminator signal in accordance with Eq.~(\ref{Eq:errorsignal}) above. We choose $\Delta_\mathrm{hop}=2.51/2\tau$. Given these specifications, the Rabi frequency $\Omega=\pi/2\tau$ renders an ideal $\pi$-pulse, and $\Delta_\mathrm{hop}$ equals the half-width-half-maximum of the corresponding Rabi spectral line. Table~\ref{Tab:sequence} gives the solitary argument $\bm{\phi}$.

For the case of Rabi spectroscopy, the clock error is simply equal to the uncompensated frequency shift, $\Delta_\mathrm{shift}-\Delta_\mathrm{step}$. This result is indicated by a straight line in Fig.~\ref{Fig:contourstandard}. For HRS, on the other hand, the clock error has cubic dependence on $\Delta_\mathrm{shift}-\Delta_\mathrm{step}$, to leading order~\cite{appeaseTZW}. Consequently, HRS enjoys much greater immunity to imperfect compensation of the frequency shift. Moreover, the clock is largely insensitive to the precise value of $\Omega$, a consequence of the $\pi$-phase jumps included in the interrogation sequence~\cite{YudTaiOat10}. Figure~\ref{Fig:contourstandard} highlights these attributes of the HRS scheme.

Up to this point, there has been no need to specify the physical mechanism responsible for $\Delta_\mathrm{shift}$. In the remainder, we identify $\Delta_\mathrm{shift}$ with the probe Stark shift discussed in the introduction, which scales with the probe intensity. Meanwhile, $\Omega$ scales with the square-root of probe intensity, assuming a single-photon clock transition. We set $\Omega=\kappa_1\sqrt{I}$ and $\Delta_\mathrm{shift}=\kappa_2I$, where $\kappa_1$ and $\kappa_2$ are proportionality constants and $I$ is the probe intensity. Thus far, we have assumed a fixed $\Omega$ and $\Delta_\mathrm{shift}$. This is in line with previous studies, leading us to familiar results such as the cubic dependence of the clock error on the uncompensated frequency shift, shown in Fig.~\ref{Fig:contourstandard}. In practice, however, fluctuations of the probe intensity occur on some level. While Fig.~\ref{Fig:contourstandard} suggests that a hyper-Ramsey clock will be largely insensitive to probe intensity fluctuations (i.e., fluctuations in both $\Omega$ and $\Delta_\mathrm{shift}$), only approximate quantitative information can be gleaned from these contours that were obtained under the assumption of fixed $\Omega$ and $\Delta_\mathrm{shift}$. This motivates us to extend the theory above to consider the potential ramifications of probe intensity fluctuations.

\begin{figure}[tb]
\includegraphics[width=0.95\linewidth]{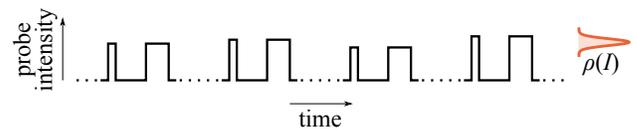}
\caption{Probe intensity $I$ versus time for HRS clock operation. Dotted lines separate independent interrogation sequences (shots) and also mark the $I=0$ level. $I$ is assumed to be constant over the extent of a given shot, with the exception of free evolution (where $I=0$). We assume that $I$ exhibits uncorrelated shot-to-shot fluctuations, described by a probability distribution $\rho\left(I\right)$. Practical clock operation may require combining spectroscopic measurements under different conditions (e.g., bias magnetic field directions or spin projections) to assess or null various systematic effects. Consequently, shots corresponding to like-conditions may be well-separated compared to the total interrogation time per shot; these intervals are collapsed and replaced with the dotted lines in this illustration.}
\label{Fig:pulses}
\end{figure}

For the present study, we assume the probe intensity is constant over the duration of an interrogation sequence (excepting free evolution), while exhibiting uncorrelated shot-to-shot fluctuations. This behavior is illustrated in Fig.~\ref{Fig:pulses}. Formally, we take $I$ to be a random variable with probability distribution $\rho\left(I\right)$; for any shot, a value of $I$ is pulled from this distribution. We introduce $f\left(I\right)$ to represent the right-hand side of Eq.~(\ref{Eq:transprob}), with the $I$-dependence being attributed to $\Omega$ and $\Delta_\mathrm{shift}$ within Table~\ref{Tab:sequence}. In the limit of no fluctuations, as assumed above, the transition probability is simply $P=f\left(I\right)$. More generally, we must weight $f\left(I\right)$ by the probability of getting a particular value of $I$, such that the transition probability for an arbitrary shot becomes
\begin{equation*}
P=\int\rho\left(I\right)f\left(I\right)dI.
\end{equation*}
This connects directly to the transition probability that would be experimentally determined in the limit of an infinite number of shots (i.e., in the limit of zero statistical uncertainty from either quantum projection noise or probe intensity noise). For simplicity, we assume $\rho\left(I\right)$ is a Gaussian distribution. We quantify intensity fluctuations by percent, which is understood to specify the standard deviation of $\rho\left(I\right)$ as a percentage of its mean. Since $\Omega$ and $\Delta_\mathrm{shift}$ no longer represent fixed parameters, we now refer to their mean values $\overline{\Omega}$ and $\overline{\Delta}_\mathrm{shift}$, where
\begin{gather*}
\overline{\Omega}=\kappa_1\int\rho\left(I\right)\sqrt{I}\,dI,
\\
\overline{\Delta}_\mathrm{shift}=\kappa_2\int\rho\left(I\right)I\,dI.
\end{gather*}

\begin{figure}[tb]
\includegraphics[width=\linewidth]{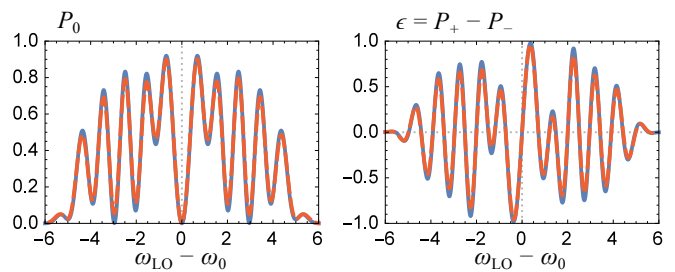}
\caption{Transition probability $P_0$ (left panel) and discriminator $\epsilon=P_+-P_-$ (right panel) for HRS with $\Delta_\mathrm{step}=\overline{\Delta}_\mathrm{shift}$ and probe intensity fluctuations as illustrated in Fig.~\ref{Fig:pulses}. The blue full curves correspond to the limit of no fluctuations, while the overlying red dashed curves correspond to 1\% fluctuations. A free evolution time $T=4\tau$, mean Rabi frequency $\overline{\Omega}=\pi/2\tau$, and mean shift $\overline{\Delta}_\mathrm{shift}=18/\tau$ are assumed for these plots. The abscissas are in units of $1/\tau$.}
\label{Fig:spectrafluc}
\end{figure}

Figure~\ref{Fig:spectrafluc} displays the transition probability $P_0$ and the discriminator $\epsilon=P_+-P_-$ under the condition $\Delta_\mathrm{step}=\overline{\Delta}_\mathrm{shift}$. The blue full curves correspond to the limit of no probe intensity fluctuations and are simply a rerendering of the curves from Fig.~\ref{Fig:spectra} above. The overlying red dashed curves correspond to finite probe intensity fluctuations. Given the conditions specified in the caption, the effect of the fluctuations is largely indiscernible on the scale of these plots. An experimental trace of the transition probability $P_0$, for example, would not produce an immediate signature of the fluctuations (e.g., significant contrast loss). Nevertheless, the presence of fluctuations alters $P_0$ and $\epsilon$ on a finer scale, which may be relevant for clock accuracy. For $\epsilon$ in particular, the full curve (no fluctuations) has a zero-crossing at precisely zero detuning, as discussed above. For the dashed curve (1\% fluctuations), this zero-crossing is displaced to $\omega_\mathrm{LO}-\omega_0=-6.1\times10^{-5}/\tau$. Thus, in the presence of probe intensity fluctuations, the condition $\Delta_\mathrm{step}=\overline{\Delta}_\mathrm{shift}$ does not identify with error-free clock operation.

\begin{figure}[tb]
\includegraphics[width=0.75\linewidth]{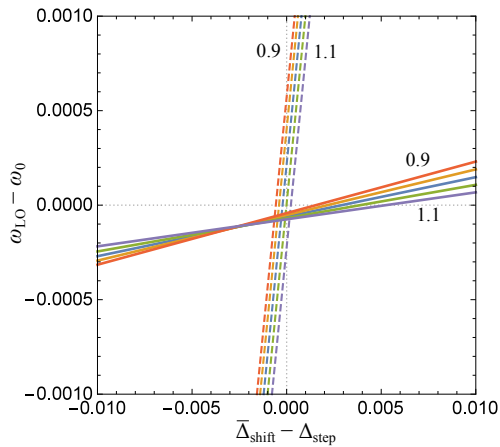}
\caption{Contours of the discriminator $\epsilon$ corresponding to a null value, $\epsilon=0$, with probe intensity fluctuations as illustrated in Fig.~\ref{Fig:pulses}. Contours are shown for both HRS (full) and Rabi spectroscopy (dashed), assuming 1\% probe intensity fluctuations. A free-evolution time $T=4\tau$ (HRS only), mean frequency shift $\overline{\Delta}_\mathrm{shift}=18/\tau$, and mean Rabi frequency $\overline{\Omega}=q\pi/2\tau$ are assumed for these plots, where $q$ is taken from 0.9 to 1.1 in increments of 0.05. Both axes are in units of $1/\tau$.}
\label{Fig:contoursfluc}
\end{figure}

Figure~\ref{Fig:contoursfluc} shows the clock error versus $\overline{\Delta}_\mathrm{shift}-\Delta_\mathrm{step}$ for finite probe intensity fluctuations. This figure is analogous to Fig.~\ref{Fig:contourstandard} above, but on a finer scale to highlight the consequences of the fluctuations. We recall that in the limit of no fluctuations, the clock error is cubic in $\overline{\Delta}_\mathrm{shift}-\Delta_\mathrm{step}$, to leading order. This is the signature feature of the HRS scheme (on the scale of Fig.~\ref{Fig:contoursfluc}, the HRS contours from Fig.~\ref{Fig:contourstandard} are essentially indistinguishable from flat lines at $\omega_\mathrm{LO}-\omega_0=0$). In the presence of the fluctuations, we see that this signature feature is lost. In particular, the clock error has both an offset (already noted in the previous paragraph) as well as a linear term in $\overline{\Delta}_\mathrm{shift}-\Delta_\mathrm{step}$. While the clock error can still, in principle, be nulled by operating with a choice value of $\Delta_\mathrm{step}$, there now exists linear sensitivity to deviations from this specific choice. Moreover, the specific $\Delta_\mathrm{step}$ that nulls the clock error is seen to depend strongly on the mean Rabi frequency $\overline{\Omega}$.

Figure~\ref{Fig:contoursfluc} also presents the clock error for Rabi spectroscopy. We recall that in the limit of no fluctuations, the clock error is given simply by $\overline{\Delta}_\mathrm{shift}-\Delta_\mathrm{step}$. This holds in the case of fluctuations, but with the inclusion of an $\overline{\Omega}$-dependent offset. This offset could have implications for clocks based exclusively on Rabi spectroscopy. For example, one strategy for handling the probe Stark shift is to operate the clock at different probe intensity levels and extrapolate to the case of zero intensity. Neglect of this offset could introduce error in this extrapolation procedure.


Thus far, we have focused primarily on the qualitative consequences of probe intensity fluctuations. To be more quantitative, we consider the Yb$^+$ single-ion clock described in Ref.~\cite{HunSanLip16}. This clock employs HRS with $\tau=30.5$~ms and $T=122$~ms. From Ref.~\cite{HunLipOkh12}, we estimate an accompanying mean frequency shift $\overline{\Delta}_\mathrm{shift}/2\pi=95$~Hz. Under these conditions and further assuming a mean Rabi frequency $\overline{\Omega}/2\pi=8.20$ Hz, we determine the tolerance in $\Delta_\mathrm{step}/2\pi$ (i.e., ``spread'' in the vicinity 95~Hz) for which the fractional clock error remains below $1\times10^{-18}$. In the limit of no probe intensity fluctuations, we find a tolerance of $818$ mHz. With just 1\% probe intensity fluctuations, on the other hand, this tolerance is shrunk by more than an order-of-magnitude to just $61$ mHz. This exemplifies how leniency in the choice of $\Delta_\mathrm{step}$ rapidly deteriorates with probe intensity fluctuations.

\begin{figure}[tb]
\includegraphics[width=0.75\linewidth]{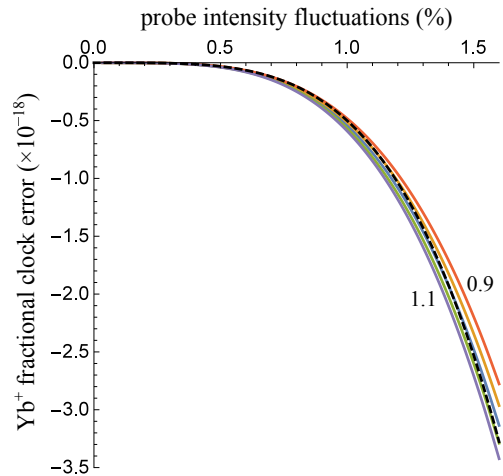}
\caption{Fractional clock error versus probe intensity fluctuations for the Yb$^+$ ion clock described in Ref.~\cite{HunSanLip16} ($\tau=30.5$~ms, $T=122$~ms, $\overline{\Delta}_\mathrm{shift}/2\pi=95$~Hz). A mean Rabi frequency $\overline{\Omega}/2\pi=q\times8.20$~Hz is assumed, where $q$ is taken from 0.9 to 1.1 in increments of 0.05. The black dashed line plots the curve $\left(-5\times10^{-19}\right)p^4$, where $p$ is the fluctuations in percent.}
\label{Fig:clockerror}
\end{figure}

To define an operational value for $\Delta_\mathrm{step}$, the Yb$^+$ clock of Ref.~\cite{HunSanLip16} further incorporates Rabi spectroscopy. In essence, $\omega_\mathrm{LO}$ is steered to null the HRS discriminator signal, while $\Delta_\mathrm{step}$ is steered to null the Rabi discriminator signal. Ideally, the probe intensity is fixed and common for the HRS and Rabi interrogations. In this case, achieving concurrent null discriminator signals implies $\omega_\mathrm{LO}=\omega_0$ and $\Delta_\mathrm{step}=\Delta_\mathrm{shift}$, independent of $\Omega$. This operating point is identified with the intersection of HRS and Rabi contours in Fig.~\ref{Fig:contourstandard}. In the presence of probe intensity fluctuations, on the other hand, achieving concurrent null discriminator signals does not imply, in particular, $\omega_\mathrm{LO}=\omega_0$. This is evident in Fig.~\ref{Fig:contoursfluc}, where the HRS and Rabi contours are seen to intersect at some point $\omega_\mathrm{LO}\neq\omega_0$. For a given level of probe intensity fluctuations, we use this intersection point to quantify a corresponding clock error.
Figure~\ref{Fig:clockerror} displays the fractional clock error versus probe intensity fluctuations. For $1\%$ fluctuations, the clock error is found to be $\approx\! -5\times10^{-19}$ and is largely insensitive to the precise value of $\overline{\Omega}$. This can be compared to the probe Stark uncertainty of $1.1\times10^{-18}$ and the total clock uncertainty of $3.2\times10^{-18}$ reported in Ref.~\cite{HunSanLip16}. The quoted probe Stark uncertainty is primarily attributed to slow drifts of $\Delta_\mathrm{shift}$, together with the servo response time for $\Delta_\mathrm{step}$. This is distinct from the error considered in this work, attributed to uncorrelated shot-to-shot fluctuations in the probe intensity. Prior works~\cite{HunLipOkh12,HunOkhLip12} suggest that the probe intensity is likely controlled at 1\% or better. The black dashed line in Fig.~\ref{Fig:clockerror} approximates the clock error with the simple function $\left(-5\times10^{-19}\right)p^4$, where $p$ is the fluctuations in percent. Because of the observed quartic behavior, the clock error drops to negligible levels for fluctuations even moderately below 1\%. At the same time, for fluctuations greater than 1\%, the clock error quickly becomes significant. With fluctuations of just 1.5\%, for instance, the clock error becomes comparable to the total clock uncertainty. 


In conclusion, we have examined the effect of probe intensity fluctuations on hyper-Ramsey spectroscopy. Such fluctuations induce correlated fluctuations in the Rabi frequency ($\Omega\propto\sqrt{I}$) and the frequency shift to the atomic transition ($\Delta_\mathrm{shift}\propto I$), both quantities that determine the spectroscopic signal. We employed a simple model for the intensity fluctuations and found that the signature robustness of the HRS scheme can be compromised, with the clock error acquiring lower-order dependence on $\overline{\Delta}_\mathrm{shift}-\Delta_\mathrm{step}$. Using the Yb$^+$ single-ion clock described in Ref.~\cite{HunSanLip16} as a quantitative example, the clock error is found to have a steep (quartic) dependence on the fluctuation level and is metrologically relevant with just $\sim\!1\%$ fluctuations. For specific applications, more complicated intensity noise models may be required, e.g., including intensity fluctuations over the duration of the interrogation sequence. In any case, the present work elucidates the general necessity to consider probe intensity fluctuations in HRS or other spectroscopic protocols~\cite{HunOkhLip12,SanHunLan17,YudTaiBas17arXiv} aimed at suppressing large probe Stark shifts.

The author thanks E.{} A.{} Donley and A.{} D.{} Ludlow for their careful reading of the manuscript. This work was supported by the National Institute of Standards and Technology/Physical Measurement Laboratory, an agency of the U.S.{} government, and is not subject to U.S.{} copyright.


\begin{thebibliography}{17}
\bibitem [{myf()}]{myfootnote}%
  \BibitemOpen
  \href@noop {} {}\bibinfo {note} {The original HRS proposal has steps 3 and 5
  interchanged~\cite{YudTaiOat10}; the variant here is mathematically
  equivalent to experimental implementations of
  Refs.~\cite{HunLipOkh12,HunSanLip16}.}\BibitemShut {Stop}%
\end{thebibliography}

%

\end{document}